# Adaptive Event Detection for Representative Load Signature Extraction

Lei Yan, Wei Tian, Jiayu Han, Zuyi Li

*Abstract*— Event detection is the first step in event-based non-intrusive load monitoring (NILM) and it can provide useful transient information to identify appliances. However, existing event detection methods with fixed parameters may fail in case of unpredictable and complicated residential load changes such as high fluctuation, long transition, and near simultaneity. This paper proposes a dynamic time-window approach to deal with these highly complex load variations. Specifically, a window with adaptive margins, multi-timescale window screening, and adaptive threshold (WAMMA) method is proposed to detect events in aggregated home appliance load data with high sampling rate (>1Hz). The proposed method accurately captures the transient process by adaptively tuning parameters including window width, margin width, and change threshold. Furthermore, representative transient and steady-state load signatures are extracted and, for the first time, quantified from transient and steady periods segmented by detected events. Case studies on a 20Hz dataset, the 50Hz LIFTED dataset, and the 60Hz BLUED dataset show that the proposed method can robustly outperform other state-of-art event detection methods. This paper also shows that the extracted load signatures can improve NILM accuracy and help develop other applications such as load reconstruction to generate realistic load data for NILM research.

## I. INTRODUCTION

DETAILED electricity consumption of residential appliance loads could help electric utilities and system operators analyze dynamic power consumption behavior of customers. It is crucial for the development of smart grid technology such as demand response [1], load forecasting [2], and peer-to-peer energy trading [3]. Appliance-level load information could be acquired with non-intrusive load monitoring (NILM), which is an event-based method initially proposed by Hart [4] in 1992. However, most of the studies have been using non-event-based NILM [5] due to the lack of high-resolution dataset.

With the development of Advanced Smart Meter, easy acquisition of high-resolution data makes it possible to conduct more studies on event-based NILM. Event detection is to find out the occurrence of state transition such as appliance turn-on/off, speed adjustment, and mode changes. Major challenges are from concentrated use (e.g., in the morning and evening) and diversity of appliances, and include high fluctuations, long transition, and near simultaneity, among others. Specifically, high fluctuation refers to the standard deviation of steady-state power larger than threshold. Near simultaneity refers to two adjacent events occurring in short time interval. For example, out of the 15 appliances in LIFTED [6], three appliances, i.e., washing machine, hair dryer, and vacuum, present high fluctuation in steady periods; eight appliances such as blender, hair dryer, kettle and refrigerator usually work in the morning. The concentrated use of appliances makes event overlap a high possibility. Long transition refers to one event lasting more than the preset window width. The transient process could be long especially when the electrical load is far larger than its rated power. For example, when the weight of clothes in washing machine is more than its specified weight, its startup transient will be longer than a normal transient. It is hard to deal with such complicated and variable measurements using event detection methods with fixed parameters. That is why most of literatures fail to detect events with high accuracy.

The purpose of event detection is to better identify appliances using load signatures that can be extracted from transient and steady periods segmented by events. Transient and steady-state load signatures are the essence of NILM that provides characteristic label for each appliance. Each load signature can be modelled as a measurable parameter giving information on the operating cycles of individual appliances. Quantitative analysis of transient signatures and their applications in NILM is currently lacking and worth further studying.

This paper proposes a dynamic time-window based method for event detection, which dynamically adjusts the size of time windows and other parameters to deal with diverse load variations. It further extracts representative load signatures based on event detection results for NILM and other applications. The proposed method is termed as window with adaptive margins, multi-timescale window screening, and adaptive threshold, or WAMMA for convenience of description. The main contributions are listed as follows:

- The proposed WAMMA method adopts window with adaptive margins as the base algorithm for event detection and applies adaptive threshold updating and macro/micro-timescale window screening to deal with unpredictable and complicated load changes including high fluctuation, long transition, and near simultaneity, respectively. Extensive case studies on three different datasets demonstrate that WAMMA can guarantee the accuracy and robustness of event detection even when using one arbitrary set of initial parameters. In comparison, existing fixed-parameter methods are less accurate and robust as they may fail to deal with those highly complex load variations.
- The proposed WAMMA method quantitatively studies transient signatures, which is first of its kind to the best of our knowledge. It detects overall transient period rather than just a change-point as in existing event detection methods. Accordingly, the detected events can divide the entire load time series data into transient and steady periods from which representative load signatures, i.e., transient and steady-state signatures, are extracted and quantified.

- This paper demonstrates that the extracted transient and steady-state signatures can help improve the accuracy and efficiency of NILM when used together and serialized using a sequential tree structure, and can help develop other applications such as load reconstruction of individual appliances, which can be used to generate realistic load data for NILM research.

The rest of this paper is organized as follows. Section II introduces the related work of event detection and load signature. Section III introduces the proposed WAMMA event detection method and load signature analysis. Simulations and discussions of results are presented in Section IV. Section V concludes this paper.

## II. RELATED WORK

### A. Review of Event Detection

Early event detection methods such as that in [4] use transient-passing step-change method to segment the power series into transient periods and steady periods in which the input does not vary by more than a preset tolerance (e.g., 15 W). Ref. [7] uses a rule-based approach to detect a preliminary set of possible start and end events for some appliances. An aggregate score associated with events based on rules is used to confirm or refute the occurrence of an event. Ref. [8] proposes a window with margin (WM) method running over the overall power measurements and calculates averages of values within two margins. It determines whether an event occurs or not by comparing the difference of the two averages and pre-specified threshold. These methods take power change and specific threshold as parameters and apply window-based or rule-based methods to determine whether an event occurs. One deficiency of a fixed-parameter method is that it lacks universality in complicated data and it has difficulty in solving the high fluctuation problem. Even though a fixed-parameter method does not have high detection accuracy, it inspires us to develop the adaptive multi-timescale window with margins method.

Besides, some methods aim to identify the specific time instance when an abrupt change occurs, namely change-point detection. Several techniques have been developed to find out these change points, which are generally termed as statistical methods such as generalized likelihood ratio (GLR) [9], log likelihood ratio detector with maxima (LLD-Max) [10], and cumulative sum (CUSUM) [11]. The statistical methods are also limited in selection of parameters. For example, the LLD-Max method uses change of mean values to calculate the log likelihood ratio when the difference of two means is beyond threshold. The two means are from pre-set windows located before and after the current time. Due to the fixed size of windows, some false alarms might occur in the long transition if the window size is smaller than the duration of transition.

Some literatures correlate known appliances transient signals with aggregated consumption signal to find the corresponding appliances, termed as matched filter (MF) method. Ref. [12] is the first one to match startup transients, which was further developed in [13] by applying Cepstrum analysis to the power signal. However, the MF method may fail as high fluctuations may mask the events of small appliances.

Ref. [14] develops a hybrid method to detect events, where two auxiliary algorithms are used to remove/reduce false alarms caused by long transitions and high fluctuations. However, its performance is still limited due to the use of fixed parameters. Table I summarizes the main characteristics of the related work. The limitations of the above-mentioned methods consist of missing an unsupervised approach for event detection, lack of flexible parameters, not dealing with near-simultaneous events (DNSE), and not extracting load signatures (LS).

### B. Review of Load Signature

Similar to images and human speech, each electric appliance has its unique load signatures. These load signatures can be from time series of voltage, current, power, and/or reactive power. Some load signatures are easily masked in variable composite load and deserve careful analysis. A partial taxonomy of load signatures is presented in Fig. 1, which includes most widely used load signatures in existing literatures. Load signatures are roughly divided into transient signatures (the green part in Fig. 1) and steady-state signatures (the yellow part in Fig. 1). Timestamp could provide extra information about the use pattern of appliances associated with a user's behavior.

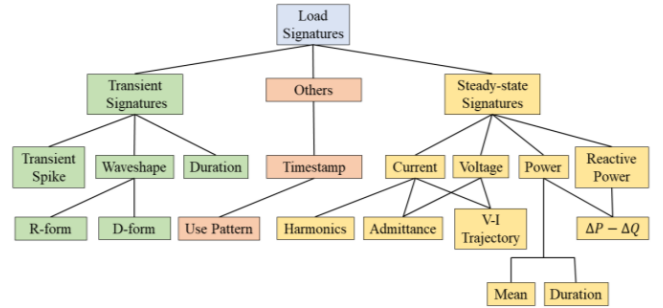

Fig. 1 A partial taxonomy of load signatures used in existing literatures.

TABLE I
SUMMARY OF EVENT DETECTION METHODS

| Approach | Characteristics | DNSE | Unsupervised | Adaptive Parameter | LS Extraction |
|---|---|---|---|---|---|
| WAMMA method (this paper) | Adaptive parameters (margin and threshold) + Multi-timescale window screening | ✓ | ✓ | ✓ | ✓ |
| Step-change [4] | Fixed window + Fixed threshold | ✗ | ✓ | ✗ | ✗ |
| Rule-based [7] | Rules | ✗ | ✓ | ✗ | ✗ |
| WM method [8] | Window with margins + Fixed parameters | ✗ | ✓ | ✗ | ✗ |
| GLR [9] | Likelihood ratio + Fixed parameters | ✗ | ✓ | ✗ | ✗ |
| CUSUM [11] | Cumulative sum + Fixed parameters | ✗ | ✓ | ✗ | ✗ |
| MF [12, 13] | Trained templates | ✗ | ✗ | ✗ | ✗ |
| Hybrid method [14] | Hybrid method + Fixed parameters | ✓ | ✓ | ✗ | ✗ |

The popularity of steady-state signature is mainly from three reasons: persistence, stability and additivity. It is a continuously stable indication of appliances' working state and additive when two or more signatures appear simultaneously. A segmented integer quadratic constraint programming (SIQCP) algorithm is proposed in [15] to solve NILM problems based on the additivity of steady-state signatures. Ref. [16] presents a unique super-state with combination of additive load states to disaggregate appliances. Ref. [17] distinguishes loads in $\Delta P$-$\Delta Q$ signature spaces, however, it fails when overlapping clusters appear.

Voltage-current (V-I) trajectory is a load signature extracted from the difference between two consecutive snapshots. Ref. [18] uses V-I trajectories to develop a taxonomy of home appliances and seven types of circuit topologies are classified with binary mapping of V-I trajectory [19]. Ref. [20] interprets V-I trajectory as weighted pixelated V-I images and uses them as input for a convolutional neural network and it is further developed in [21] using transfer learning. Ref. [22] presents the admittance of appliances, which is defined as the quotient between the current and voltage and shows that some appliances have similar admittance signatures.

As presented in [17] and [22], two types of appliances may have similar steady-state signatures; however, transients are more different and provide more deciding information than steady-state signatures. Transient signatures are classified into rectified form (R-form) and delta-form (D-form), and their difference is whether the transient spike power goes far beyond the steady-state power. Ref. [23] searches for a precise time pattern of D-form over many timescales. Ref. [24] classifies the power curves of events as triangle and rectangle and extracts five items including *starttime, peaktime, peakvalue, steadytime, steadypower* to express the transition process.

## III. Proposed Methodology

This paper proposes a dynamic time-window approach with adaptive parameters to detect event in unpredictable, complicated, and wide variety of residential loads. While existing fixed-parameter methods may fail to detect events of electrical appliances that are greatly different in transitional shapes and durations as well as steady-state power fluctuation, the proposed WAMMA method effectively solves these problems by right of its adaptability and multi-timescale screening. A sequential structure is developed as well to store load signatures according to the order of appearance.

### A. Framework of the Proposed Method

Fig. 2 illustrates the structure of the proposed WAMMA method, load signatures and their applications. The proposed method is specifically designed to solve the challenges of high fluctuation, long transition, and near simultaneity. The original window with margins method aims to reduce the effect of high fluctuation by using average of values within the margins that works as a mean filter. WAMMA adopts window with adaptive margins as the base algorithm to capture the transition process by automatically adjusting the border lines of the margins to make margins stay in steady periods. The macro-timescale window is used to detect long transitions and the micro-timescale window, which is much narrower than the initial window, is used to further detect near-simultaneous events.

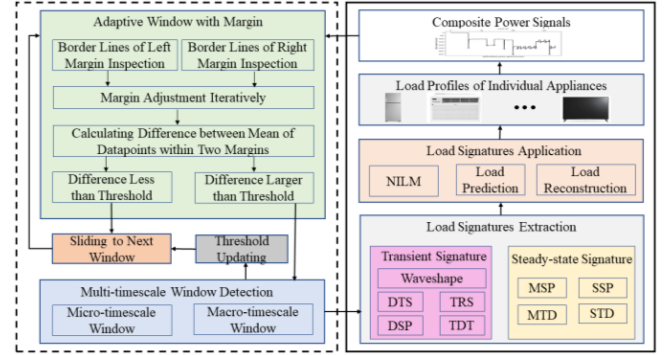

Fig. 2 Framework of WAMMA and load signature. Left plot in dashed rectangle: WAMMA; right plot in solid rectangle: load signatures.

The proposed sequential load signature tree is used for organizing chosen transient and steady-state signatures in an orderly fashion. The chosen signatures will be used for NILM and other applications such as load reconstruction. As directly collecting load data is time consuming, constructing load profiles of different types of appliances by studying their load signatures may be a more efficient method. In this way, more datasets with different data resolutions (with different signatures) can be created to help develop the relevant research.

### B. Review of Window with Margins Method

Window with margins method was firstly proposed in [7]. Fig. 3 shows the ideal working state and required parameters. It has three parameters: number of datapoints in the window ($N_w$), number of datapoints in the margins ($N_{Lm}$, $N_{Rm}$), where $N_{Lm} = N_{Rm} = N_m$ in the default condition, and preset threshold $P_{thre}$.

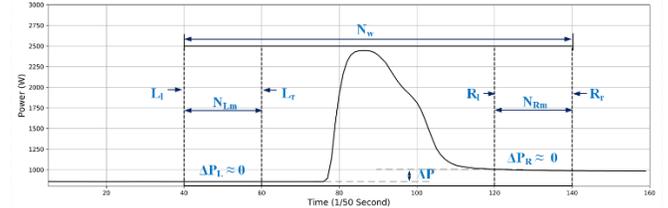

Fig. 3 Illustration of window with margin method.

Four coordinates are determined based on the datapoints being processed, including left border line coordinate value of left margin ($L_l$), right border line coordinate value of left margin ($L_r$), left border line coordinate value of right margin ($R_l$), and right border line coordinate value of right margin ($R_r$). $\Delta P_L$ and $\Delta P_R$ in Eq. (1) are absolute differences of observations on the margin border lines. $\Delta P$ is the difference between means of datapoints in left margin ($\mu_l$) and mean of datapoints in right margin ($\mu_r$) as shown in Eqs. (2) and (3).

$$\Delta P_L = |o_{L_r} - o_{L_l}|, \Delta P_R = |o_{R_r} - o_{R_l}| \quad (1)$$

$$\mu_l = \frac{1}{N_{Lm}} \left( \sum_{i=L_l}^{L_r} o_i \right), \mu_r = \frac{1}{N_{Rm}} \left( \sum_{i=R_l}^{R_r} o_i \right) \quad (2)$$

$$\Delta P = \mu_r - \mu_l \quad (3)$$

where $o_i$ refers to the $i_{th}$ observation. When $\Delta P$ is beyond the threshold value $P_{thre}$ (i.e., $\Delta P > P_{thre}$ or $\Delta P < -P_{thre}$), an event is said to have occurred.

## C. Adaptive Window to Capture Transients

The main drawback of the original window with margins method is its use of fixed parameters that hinders the precise capture of different events. A too narrow window may falsely identify a long transition as multiple events, while a too wide window may include multiple events. Some problems with fixed parameters are shown in Fig. 4. As the window runs over continuous data stream, there will be different kinds of situations where the left and right margins rest on transitions. The margin could either take up half of the transition like (a) and (b) or ride on both sides of a transition such as (c) and (d) in Fig. 4. The proposed WAMMA method will remedy these problems by adaptively adjusting the widow margins based on the data being processed. Such dynamic time-window method is effective to deal with complicated time series in practice.

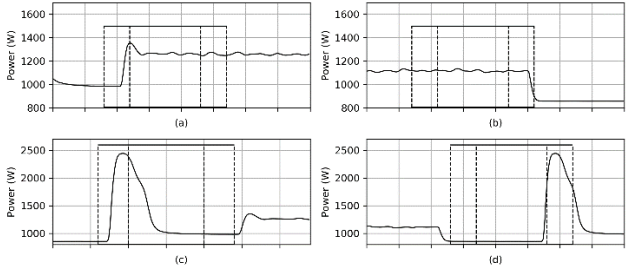

Fig. 4 Cases of margins resting on transitions.

As the objective of the window with margins method is to make the margins stay on steady periods as shown in Fig. 3, a natural extension is to adjust the border lines of margins and window widths to accommodate transitions as shown in Fig. 5. $L_l$ keeps fixed because it is just the $R_r$ of previous window and stays on steady period. $L_r$ will shift toward left in cases like Fig. 4 (a) and (c). $R_l$ and $R_r$ will move toward right until both are on steady period. The way to judge whether the $L_l$ and $L_r$, $R_l$ and $R_r$ are on steady periods is to check the difference of the observation as shown in Eq. (1). If $\Delta P_L$ or $\Delta P_R$ is larger than the specified tolerance, the associated border lines will adjust. However, only small $\Delta P_L$ and $\Delta P_R$ cannot guarantee that the margins are on steady periods. When the observations on the left and right borders of margins do not have significant difference such as (d) in Fig. 4, an event is still likely to occur. Threshold inspection is necessary but not sufficient, and the change trend and range of transition need to be further checked.

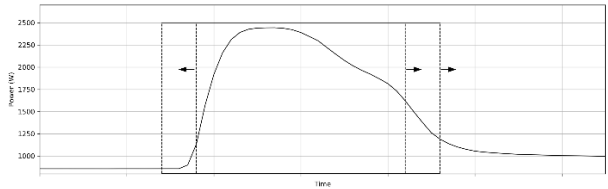

Fig. 5 Illustration of border line moving direction

The traditional CUSUM [25] calculates the mean of datapoints within a window and then computes cumulative sum. It does not illustrate the relationship between consecutive datapoints because an event generally occurs when the sum of changes among consecutive datapoints is larger than the threshold. This paper proposes a modified CUSUM to check whether an event truly occurs. The difference between consecutive datapoints is shown in Eq. (4) which can be used for calculating cumulative summation and changing trend. As shown in Eq. (5), if the cumulative sum $S_i$ is larger than $P_{thre}$, an event is supposed to occur, where $S_i = S_{i-1} + d_i$, and $E_i=1$ indicates that an event occurs at time index i; otherwise, $E_i=0$.

$$d_i = o_i - o_{i-1} \tag{4}$$

$$E_i \sim \underset{i}{\mathrm{argmax}}(P_{thre}, S_i) \tag{5}$$

Changing trend excludes the possibility of regarding high fluctuation as true events. A high fluctuation may cause a high $S_i$, but its continuous oscillation results in frequent direction change of waveshape trend. In such case, the change sign of consecutive datapoints, i.e., $\mathrm{sgn}(d_i)$ in Eq. (6), alternates between positive (1) and negative (-1). This information can be used to remove the false alarms caused by high fluctuation. In addition, parts of transitions may go up or down slowly without obvious variation so that threshold alone cannot determine whether the border line should move or not. Change trend will guide window adjustment instead. In this paper, if the negative change signs within a window account for more than 60%, the right margin will move rightward even though the change value is smaller than the threshold. Such movement will stop until the conditions for both change signs and change values are met. After adaptive adjustment, all the border lines in Fig. 4 will move to steady period as shown in Fig. 6.

$$\mathrm{sgn}(d_i) = \begin{cases} -1 & \text{if } d_i < 0 \\ 0 & \text{if } d_i = 0 \\ 1 & \text{if } d_i > 0 \end{cases} \tag{6}$$

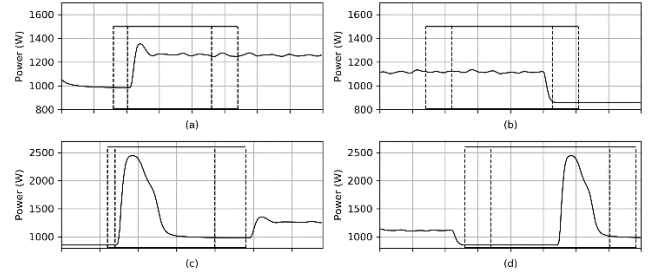

Fig. 6 Adaptive adjustment of border lines in Fig. 4.

## D. Multi-timescale Window Screening

Adjustment of window with margins guarantees that the margins can accommodate steady periods. However, it is likely that a steady period is a saddle period of a long transition as shown in Fig. 7, or there are more than one appliance changing states at the same time. The issues of long transition and near simultaneity along the time dimension have not been solved completely before. Thus, the method of multi-timescale window screening (including macro-timescale screening and macro-timescale window screening) is proposed to detect if near-simultaneous events or long transitions occur.

Macro-timescale window screening aim to detect long transition by using two components. The first is one margin attempt which uses one more margin to inspect the change trend and range of the following observations. In the example shown in Fig. 7, the long transition is longer than 20 seconds (10 times the original default window size) and the right margin stops after a number of rightward shifts. However, the right margin is still on saddle period of a long transition that has not finished yet. The one margin attempt shown in grey color finds out that more than 60% of change signs are negative, so the right margin continues shifting rightward. The second one is concatenation of near-simultaneous events. If the first event is classified to some appliance in a low probability in the following appliance identification step, it is very likely that two events are from the same long transition. In this case, the concatenated event will be inferred again to find out which appliance is changing state. The second component works well in conjunction with load disaggregation in a later stage. As this paper is mainly about event detection, it is not considered in the case study.

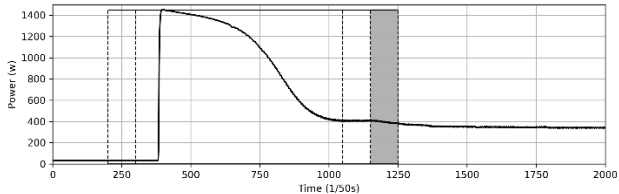

Fig. 7 Illustration of one margin attempt of macro-timescale window

Micro-timescale window screening aims to distinguish close events in a finer timescale by using change trend and range. It works after the macro-timescale window screening completes. A small window that has the same width as the left margin scans from left to right in the window to detect suspicious events. If there is a suspicious event, the NILM system in the load disaggregation stage will infer whether two adjacent events occur based on load signatures of individual appliances. However, it is limited in two cases. The first one is high fluctuation as it is easy to bring false alarms in the fine details; the other one is event overlap when two events with small time difference would be detected as one.

Using micro-timescale window, it is hard to know whether the detected event is part of a long transition or overlapping of two different events. Therefore, event detection does not work independently in the entire NILM system but cooperates with load signatures to identify appliances. If there are two adjacent events, NILM will classify the first event to some appliance. If it fails, the whole transition will be considered as one event. Micro-timescale window screening guarantees not to miss any possible events in finer scale and load signatures ensure the true event detection and improve the accuracy of NILM. Since this paper focuses more on the event detection, it will count each detected event as one rather than consider connecting events.

### E. Threshold Updating

If no event is detected, the mean and standard deviation of the datapoints being processed will be calculated. The standard deviation reflects the degree of fluctuation of datapoints. The threshold decides whether an event is occurring or not, which will fail to work when the fluctuation thus the standard deviation is high. To remedy the effect of high oscillation, we propose that the threshold should be updated based on the datapoints being processed. If the standard deviation of the datapoints is held within a small range, the threshold needs not be updated. Conversely, the threshold value will be updated.

Furthermore, threshold update should be primarily based on appliances characteristics. In this paper, washing machine has the highest variation of power consumption among all appliances tested. To reduce its effect, when 20% of the calculated standard deviation is bigger than the initial threshold, the threshold will be updated to 20% of the standard deviation.

### F. Load Signatures and Sequential Load Signature Tree

Load signatures are measurable parameters that can provide information about operating cycles of individual appliances. However, to the best of our knowledge, there has not been other literatures to build quantitative model for transient signatures and integrate them into the NILM formulation. The model in this paper tends to transform transient signatures analysis from qualitive to quantitative in the following way.

- When events are detected using WAMMA, the time series can be divided into transient and steady periods.
- After clustering the steady periods, the state number of each appliance can be determined and steady-state signatures are also extracted as statistical results.
- The transition between two different steady states is labelled as well from which transient signatures can be extracted.

It is important to note that a detected event is a process rather than just a change-point. Even though it is hard to describe the process with a mathematical model, load signatures can be used to depict the change range and trend as shown in Fig. 8. Transient signatures include difference between transient spike value and previous steady-state mean value (DTS), time to reach transient spike power (TRS), difference between steady-state mean and the previous steady-state mean (DSP), time duration of transition process (TDT). Steady-state signatures include steady-state power (SSP) and steady-state time duration (STD). All these signatures together represent the nature of appliance operating cycles and enable information processing in both power and time domains at the same time. Thus, they are called representative load signatures in this paper. It should be noted that each signature is modelled as a Gaussian distribution characterized by two parameters including mean and standard deviation.

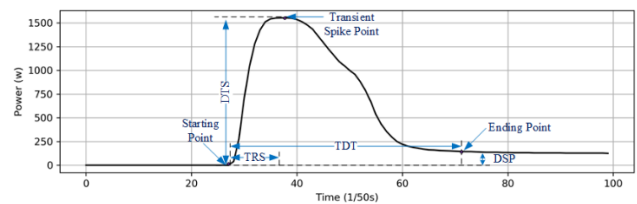

Fig. 8 Illustration of load signatures in refrigerator transition process.

To extract load signatures, it is necessary to find out the times and power values of starting point, transient spike point, and ending point in Fig. 8. Starting point refers to a datapoint whose value is close to the mean of previous steady period and the change scope of whose following points is larger than the current threshold. Transient spike point is a datapoint whose value is larger than any other neighboring datapoint. Ending point is a datapoint whose value is close to the mean of the next steady period and the change range of its following datapoints is less than the current threshold.

The proposed sequential load signature tree representation serializes the extracted signatures in the order they are stored as shown in Fig. 9. It starts from the root node, processing nodes sequentially until reaching the attribute nodes that represent the steady-state signatures. The second signature is transient waveshape, which could function as a classification layer that increases the retrieval efficiency as it divides the signatures into two categories: R-form or D-form. The next four layers are DTS, TRS, DSP and TDT that are modelled as Gaussian distribution. The sequence of these layers is listed in chronological order. The seventh layer is label indicating transition from one state to another state whose attributes are SSP and STD referred to as the eighth and ninth signatures.

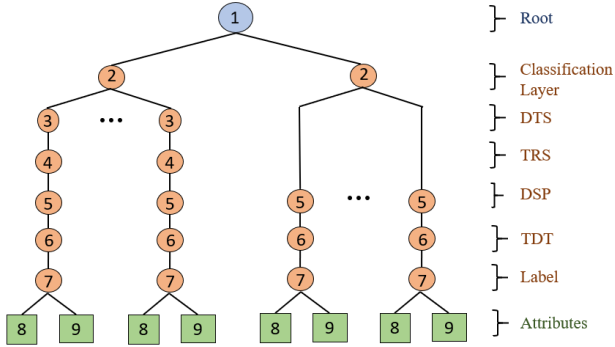

Fig. 9 Sequential load signature tree representation.

The sequential tree implementation makes it easy to retrieve load signature information along the preordered chain. The retrieved information can be used in NILM to infer which appliance is changing states in a real-time way according to the transient signatures. The inferred results will be further verified based on the steady-state attributes. It can also be used to easily reconstruct the transition and steady-state process by enumerating preordered traversal.

## IV. CASE STUDY

In this section, the proposed WAMMA method is tested on datasets of different frequencies to simulate different cases that may happen in the real world. Four cases are designed to prove its effectiveness. The first two cases are tested on a 20Hz dataset [14] to show how WAMMA works for long transition, high fluctuation, and near simultaneity and to compare WAMMA with other methods. The third case is conducted on synthetic measurements in the 50Hz LIFTED dataset [6] to prove that WAMMA with one set of preset parameters outperforms other methods with scores of parameters. The fourth case on the 60Hz BLUED public dataset [26] is to demonstrate that WAMMA robustly outperforms other methods. The application of representative load signatures is presented as well to illustrate the improvement on NILM and load reconstruction. All tests run in Python 3.6 on a desktop with a 4.2GHz intel i7-7700K CPU and 16G memory.

### A. Methods of Comparison with Different Parameters

It is not fair to compare with other methods such as step-change method [4], window with margin method [8], LLD-Max [10], CUSUM [25], and hybrid method [14] with only one set of parameters, so this paper will test the existing methods with multiple sets of parameters and use their best result to compare with the proposed WAMMA method.

The step-change method has two parameters, ratio r which is window width over sampling rate of the dataset and threshold $p_{thre}$. Window with margin method [8] has four parameters: primary window width Wd, secondary window width Wf, number of samples on each margin Nm and threshold $p_{thre}$. So, ratios rd, rf and rm are used to represent the first three parameters that are the corresponding ratios over sampling rate. Narrower window may bring some false alarms from long transition but may detect near-simultaneous events more accurately. Wider window may capture overall long transition but may regard adjacent events as one. Threshold has a great effect on event detection accuracy as well. In order not to miss small appliance events, threshold should be as small as possible. However, small threshold will bring some false alarms.

CUSUM in [25] mainly has two tunable parameters, i.e., ratio r and threshold $p_{thre}$. Ratio r is the ratio of the window width w over the sampling rate of dataset. The LLD-Max method [10] has four tunable parameters: pre- and post-event window width $w_0$ and $w_1$, threshold $p_{thre}$ and maximum precision window $M_{pre}$. For simplicity, the pre- and post-event window widths are set as the same value w. So, three parameters, i.e., w, $p_{thre}$ and $M_{pre}$, are considered to validate its effectiveness.

Parameters of GLR include the number of samples in the event detection window ($w^e$), the pre-event and post-event window sizes ($w_b^l$, $w_a^l$), and the minimum number of votes ($v_{min}$) as discussed in [27]. But the best result rather than the whole parameters space will be tested here and the same is for the hybrid method as well.

To better suit different datasets and being easy to tune parameters, the proposed WAMMA method has three tunable parameters: $r_m$, $r_w$ and $p_{thre}$, where $r_m$ and $r_w$ are ratios of $N_m$, $N_w$ and sampling rate f, respectively.

Various parameter sets constitute different parameter combinations that are indexed according to the possible values of parameter r and $p_{thre}$. For example, if the ratio and the threshold of the CUSUM method are set to {0.5, 1} and {100, 200}, respectively, there will be four combinations of CUSUM parameters as listed in Table II.

TABLE II
COMBINATION OF CUSUM PARAMETERS

| No. | Ratio r | Threshold $p_{thre}$ (w) |
|---|---|---|
| 1 | 0.5 | 100 |
| 2 | 0.5 | 200 |
| 3 | 1 | 100 |
| 4 | 1 | 200 |

The event detection results are evaluated by using true positive percentage TPP = TP/EG, false positive percentage FPP = FP/ED, and false negative percentage FNP = FN/EG, where EG is the number of ground-truth events, ED is the number of detected events; TP, FP and FN are the number of true positive, false positive, and false negative events, respectively. This paper will also use the $f_1$ score to identify the best parameter as shown in Eq. (7).

$$f_1 = \frac{TP}{TP + 0.5(FP + FN)} \quad (7)$$

### B. Performance Evaluation on 20 Hz Dataset

This dataset [14] is collected from real house, lab and office recorded at 20 Hz. The first case with 20 events including long transients and high fluctuations is shown in Fig. 10. The top plot is complete test data, and the bottom two plots show long turn-on transient of range hood and high fluctuation of hair dryer.

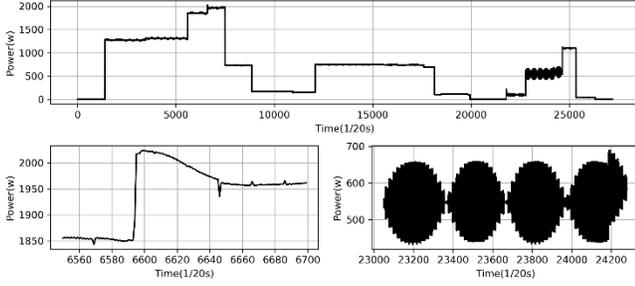

Fig. 10 Test data of 20 Hz with long transient and high fluctuation

The proposed WAMMA method will update threshold when there are high fluctuations. For high fluctuation shown in Fig. 10, the threshold is manually set to 20% of standard deviation, i.e., 20.6W, which is larger than initial parameter 15W. It means that only when the difference of mean values is larger than 20.6W, an event is considered to occur. Fig. 11 shows event detection results of WAMMA that includes all 20 true events.

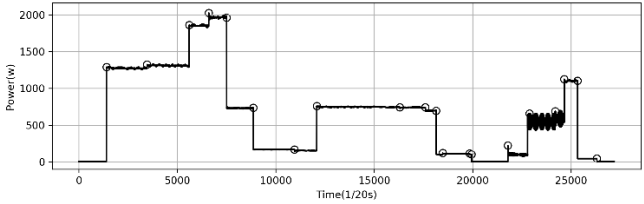

Fig. 11 Event detection results with long transient and high fluctuation

Since the action times of near-simultaneous events are so close, it is easy to regard these events as one. The second case is shown in Fig. 12, where the bottom three plots are different near-simultaneous events. In particular, the time between two events in the bottom right plot is less than 1 second, which is shorter than the default window width of 2 seconds.

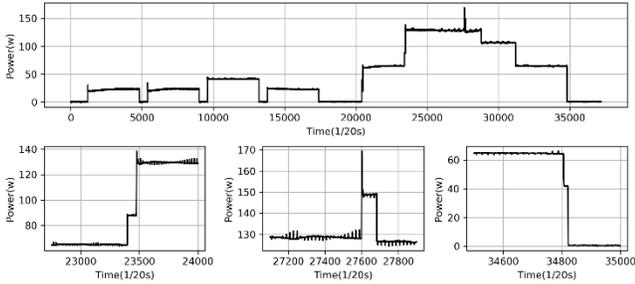

Fig. 12 Test data of 20Hz with near-simultaneous events

The proposed micro-timescale window inspects the fine-scale details to check if there are two adjacent events within short time intervals. Since the initial parameter $r_m$ is 0.2, i.e., 4 datapoints of margin width that is less than the shortest interval of two events, the proposed method is able to distinguish the two adjacent events even if they are in the same window. In comparison, other existing methods such as the one in [8] do not have the multi-timescale screening step, which may be the reason why the near-simultaneous events are always detected as one. To better compare with WAMMA, different initial parameters are shown in Tables III, IV, and V. Each combination is indexed according to the rule in Table II.

TABLE III
PARAMETER RANGES OF WINDOW WITH MARGINS

| Parameter | Min | Max | Increment |
|---|---|---|---|
| rd | 0.5 | 1.5 | 0.5 |
| rf | 0.5 | 2 | 0.5 |
| rm | 0.1 | 0.2 | 0.1 |
| $p_{thre}$ (w) | 15 | 20 | 5 |

TABLE IV PARAMETER RANGES OF CUSUM

| Parameter | Min | Max | Increment |
|---|---|---|---|
| r | 0.5 | 1.1 | 0.3 |
| $p_{thre}$ (w) | 30 | 480 | 50 |

TABLE V PARAMETER RANGES OF LLD-MAX

| Parameter | Min | Max | Increment |
|---|---|---|---|
| w | 0.3 | 0.9 | 0.3 |
| $p_{thre}$ (w) | 15 | 25 | 5 |
| $M_{pre}$ | 1 | 4 | 1 |

The best and worst results in terms of $f_1$-score including number of detected events (ED) and number of detected true events (ET) (out of 18 true events) for window with margins, CUSUM and LLD-Max methods are shown in Table VI. It can be observed that the detection results are highly affected by the parameters. This means that the initial parameters are crucial to event detection and an improper preset parameter could lead to extremely poor result. The LLD-Max method detects event point-by-point so that even though it could detect all 18 true events, it misidentifies some false events as true ones. This kind of misidentification would be worse if more long transitions and high fluctuation appear. For window-based methods, the bigger window width and threshold would miss some near-simultaneous events as in the case of window with margins and CUSUM methods, which only detect 3 events and 1 event in the worst case, respectively.

TABLE VI
THE BEST AND WORST RESULTS OF WM, CUSUM, AND LLD-MAX

| Methods | Cases | ED | ET | $f_1$-score |
|---|---|---|---|---|
| Window with margins | Best | 16 | 16 | 94.1% |
| | Worst | 3 | 3 | 28.5% |
| CUSUM | Best | 17 | 17 | 97.1% |
| | Worst | 1 | 1 | 10.5% |
| LLD-MAX | Best | 22 | 18 | 91.7% |
| | Worst | 9 | 6 | 40.0% |

Table VII compares the best results of the above three methods along with those of the hybrid method [14] and WAMMA. As different appliances have different transition process, it is hard to tune the suitable parameters such as pre- and post-event window width to achieve good accuracy. This test reveals drawbacks of existing methods, while WAMMA and the hybrid method outperform the others.

TABLE VII
PERFORMANCE COMPARISON OF DIFFERENT METHODS ON 20HZ DATASET

| Methods | TPP | FPP | FNP | f1 |
|---|---|---|---|---|
| Window with margins | 88.9% | 0% | 11.1% | 94.1% |
| CUSUM | 94.4% | 0% | 5.6% | 97.1% |
| LLD-Max | 94.4% | 18.1% | 5.6% | 91.7% |
| Hybrid method | 100% | 0% | 0% | 100% |
| **WAMMA** | 100% | 0% | 0% | 100% |

C. *Performance Evaluation on LIFTED Dataset*

LIFTED is sampled at 50Hz with motor-driven appliances that are not easy to detect with fixed parameters. The synthetic data includes 10 household appliances and 130 true events.

The initial parameters $r_m$, $r_w$, and $p_{thre}$ of the WAMMA method are set to 0.25, 2, and 15W, respectively. Fig. 13 shows the event detection results of washing machine spin mode with long transient. The top, middle, and bottom plots are the results of the full WAMMA method, the full WAMMA method without adaptive window with margins (FWA), and the full WAMMA method without macro-timescale window (FWM), respectively. While all three methods can detect all 10 events during the five cycles of spin mode, the FWA and FWM methods produce 4 and 2 false alarms, respectively.

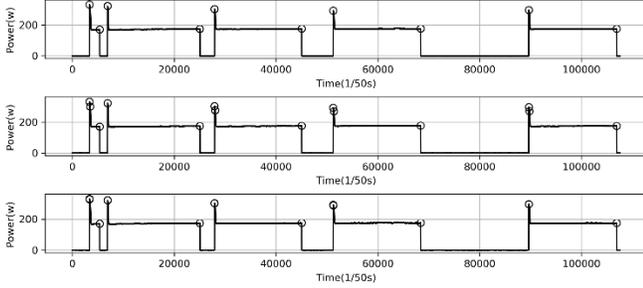

Fig. 13 Event detection results of portable washing machine with long transient. Top: full WAMMA method; middle: FWA; bottom: FWM

Fig. 14 shows the event detection results of the entire synthetic dataset. The data and associated events are on the top plot; the bottom three plots show the three missed events by WAMMA. The missed events are from the same transition of washing machine at the time of closure of water supply. The transition gradually decays rather than represent step-like change. It is hard to detect this kind of events directly. Besides, the number of detected events is 127 without any false alarms.

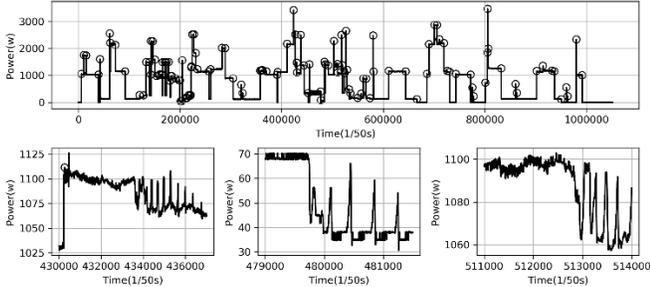

Fig. 14 Event detection results of synthetic data of LIFTED dataset. Top: all the event detection results; bottom: three missed true events.

Next, the impact of parameter selection on detection results is studied. The parameters for the window with margins, CUSUM, LLD-MAX, and WAMMA methods are listed in Tables VIII, IX, X, and XI, and corresponding performances are shown in Fig. 15. It can be observed that when the threshold is smaller, more events are detected including both true positive and false positive events. The variations of threshold can cause big change in the number of detected events. The best result with the highest $f_1$ score is achieved at $36^{th}$, $13^{th}$, and $9^{th}$ parameters for window with margins, CUSUM, and LLD-MAX, respectively. As mentioned before, more long transitions and high fluctuation cause more misidentification even though LLD-MAX can detect 129 out of the 130 true events.

For WAMMA, it can be observed that 1) the performance is robust as there is no significant difference among different parameters; 2) the accuracy is very high with TPP above 95% and FPP below 5%. It means that even if it is hard to choose the best parameter, any set of arbitrarily chosen parameters can achieve similarly good performance in actual applications.

TABLE VIII
PARAMETER RANGES OF WINDOW WITH MARGINS

| Parameter | Min | Max | Increment |
|---|---|---|---|
| rd | 1.5 | 2 | 0.5 |
| rf | 1.5 | 2.5 | 0.5 |
| rm | 0.2 | 0.4 | 0.2 |
| $p_{thre}$ (w) | 20 | 40 | 10 |

TABLE IX PARAMETER RANGES OF CUSUM

| Parameter | Min | Max | Increment |
|---|---|---|---|
| w | 0.5 | 1.5 | 0.5 |
| $p_{thre}$ (w) | 50 | 1850 | 100 |

TABLE X PARAMETER RANGES OF LLD-MAX

| Parameter | Min | Max | Increment |
|---|---|---|---|
| w | 1 | 2.5 | 0.5 |
| $p_{thre}$ (w) | 15 | 25 | 5 |
| $M_{pre}$ | 1 | 3 | 1 |

TABLE XI PARAMETER RANGES OF WAMMA

| Parameter | Min | Max | Increment |
|---|---|---|---|
| $r_m$ | 0.1 | 0.5 | 0.2 |
| $r_w$ | 2 | 3 | 0.5 |
| $p_{thre}$ | 20 | 30 | 5 |

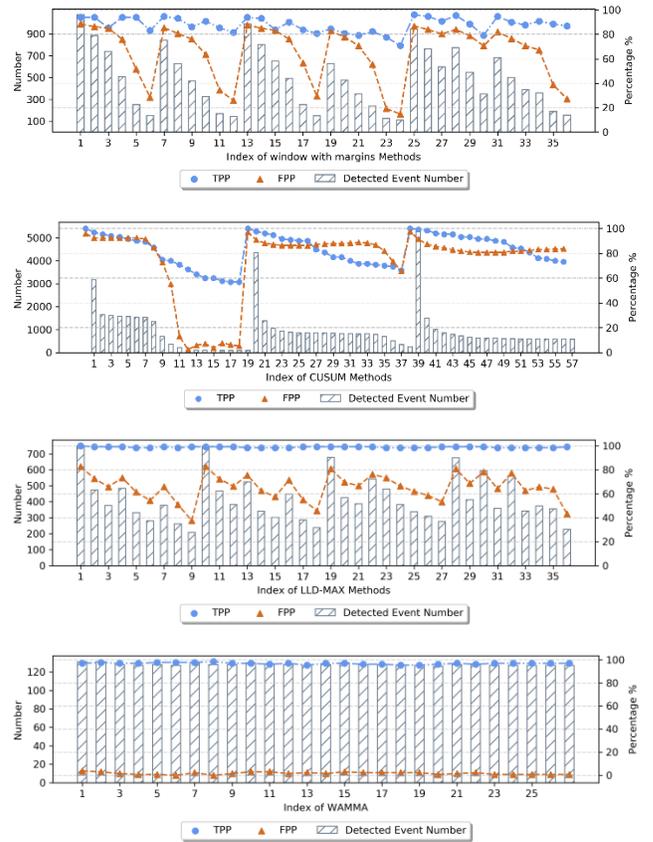

Fig. 15 Performance of window with margin, CUSUM, LLD-MAX and WAMMA methods with different parameters

TABLE XII
PERFORMANCE COMPARISON OF DIFFERENT METHODS ON LIFTED

| Methods | TPP | FPP | FNP | f1 |
|---|---|---|---|---|
| Window with margins | 86.9% | 27.1% | 13.1% | 81.2% |
| CUSUM | 66.9% | 2.3% | 33.1% | 79.1% |
| LLD-Max | 98.5% | 36.6% | 1.5% | 83.7% |
| **WAMMA** | 97.7% | 0% | 2.3% | 98.8% |
| Average of WAMMA | 96.5% | 1.7% | 3.5% | 97.3% |

The best performances of window with margins, CUSUM, LLD-Max, and WAMMA are shown in Table XII. It can be observed that WAMMA outperforms other methods by a wide margin. For example, the f1 score of WAMMA is at least 15% higher than those of other methods. In addition, WAMMA is robust with the average performance very close to the best one.

*D. Performance Evaluation on BLUED Dataset*

In the fourth case, the BLUED public dataset with 121 recorded events for 12 different appliances is used to test different methods. As shown in Fig. 15, the selection of parameters has a great impact on the detection results. To validate the robustness of WAMMA on a public dataset, 27 parameter combinations are set as in Table XIII.

TABLE XIII
PARAMETER RANGES OF WAMMA METHOD

| Parameter | Min | Max | Increment |
|---|---|---|---|
| $r_m$ | 0.1 | 0.5 | 0.2 |
| $r_w$ | 2 | 3 | 0.5 |
| $p_{thre}$ | 20 | 30 | 5 |

As in the previous case study, the results for WAMMA with different parameters are very similar. The best result is achieved with the 8th parameter combination as shown in Fig. 16 whose parameters of $r_m$, $r_w$ and $p_{thre}$ are 0.3, 2, 25, respectively. The plot on the bottom left shows the details of near-simultaneous events and the bottom right one shows the missed event. Because the time interval of these two near-simultaneous events (0.15s) is less than the width of margin (0.3s), they occur in the same margin and are detected as one event. The width of margin is the smallest timescale of the proposed method, and any two events occurring less than the width of margin may be regarded as one.

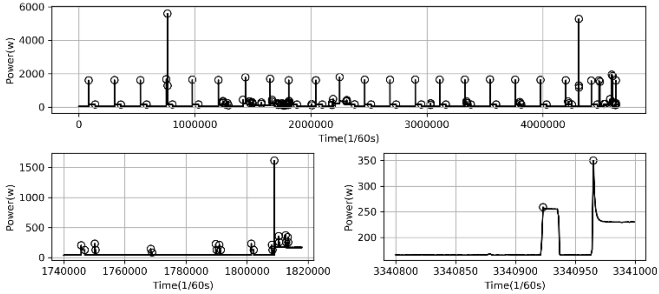

Fig. 16 Event detection results of BLUED dataset using WAMMA

Table XIV shows the performance comparison of different methods. The parameters of all the methods are tuned to achieve the highest $f_1$ score. As mentioned before, WAMMA detects 120 true events without any false alarm and misses 1 event. Even the average performance of WAMMA outperforms all other methods except the hybrid method that has an f1 score 0.2% higher. This again demonstrates the robustness of WAMMA.

TABLE XIV
PERFORMANCE COMPARISON WITH OTHER PUBLISHED WORK ON BLUED

| Methods | TPP | FPP | FNP | f1 |
|---|---|---|---|---|
| Window with margins | 93.4% | 8.1% | 6.6% | 92.7% |
| CUSUM | 95.8% | 8.6% | 4.2% | 93.7% |
| LLD-Max | 96.7% | 26.1% | 3.3% | 86.8% |
| GLR | 96.7% | 24% | 3.3% | 87.6% |
| Hybrid method | 96.7% | 0.81% | 3.3% | 97.6% |
| Step-change method | 97.5% | 61.7% | 2.5% | 75.2% |
| **WAMMA** | **99.2%** | **0.8%** | **0.8%** | **99.2%** |
| Average of WAMMA | 97.1% | 2.2% | 2.9% | 97.4% |

It should be noted that it is not always possible to tune the parameters to the best performance. Therefore, the initial parameters have a great impact on the detection results. This is a great advantage of WAMMA compared with others.

*E. Load Signatures Analysis*

Compared with other methods, another advantage of the proposed WAMMA method is to capture the whole transition process rather than just a change-point. The detected events segment the measurements into transient and steady periods from which transient and steady-state signatures can be extracted. The signatures are serialized as a sequential tree representation that can be used for load monitoring and reconstruction. A complete discussion the applications to load monitoring and reconstruction is out of the scope of this paper and will be the subject in another paper. In this section, we will take kettle and vacuum as an example to illustrate how it works.

The measurable transient signatures of active power are DTS $\alpha$, DSP $\beta$, TRS $\gamma$, and TDT $\delta$, and the steady-state signatures are SSP $\mu$ and STD $\tau$. The signatures are modelled as Gaussian distribution with parameters shown in Table XV where the first value is mean and the second one is standard deviation.

TABLE XV LOAD SIGNATURES OF KETTLE AND VACUUM

| | $\alpha$ (w) | $\beta$ (w) | $\gamma$ (s) | $\delta$ (s) | $\mu$ (w) | $\tau$ (s) |
|---|---|---|---|---|---|---|
| Kettle | (1139, 9.8) | (1138, 10.1) | (0.48, 0.28) | (0.48, 0.28) | (1027, 5.2) | (514, 43.2) |
| Vacuum | (2339, 71) | (1101, 64) | (0.14, 0.1) | (1.14, 0.33) | (1002, 22.1) | (225, 16.5) |

The first application of the extracted load signatures is load monitoring that could be used to identify different kinds of appliances. Most of existing methods such as those in [5-6] and [15-16] used steady-state signatures in NILM. It may be hard to distinguish kettle and vacuum only depending on the steady-state power as shown in top left plot of Fig. 17, where its distribution highly overlaps. If transient signatures DTS, DSP, and TDT are considered, load signatures of these two appliances are greatly different and easy to recognize as shown in other plots of Fig. 17. If all the load signatures are integrated in one model, it will greatly increase identification accuracy.

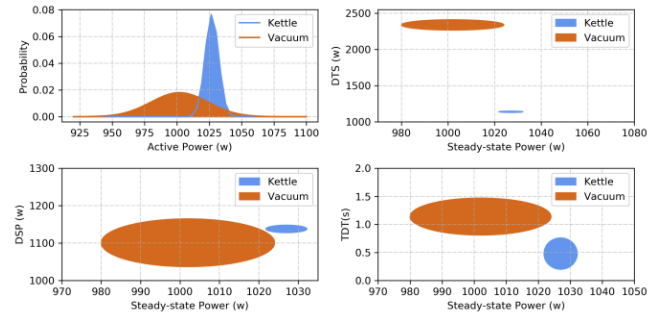

Fig. 17 Load signatures representation of kettle and vacuum in LIFTED. Top left: distribution of steady-state power; top right: 2-D representation of DTS and SSP; bottom left: 2-D representation of DSP and SSP; bottom right: 2-D representation of TDT and SSP.

Another application of load signatures is to reconstruct the operating data of each appliance. Fig. 18 shows one cycle of ground truth and three constructed loads by fitting the transition and steady process based on the sequential signature tree for kettle and vacuum. Even though the reconstructed loads are different from the original load, the signatures still follow the same Gaussian distribution and reflect the cases in practice. Using the representative load signatures, it is simple to build numerous synthetic loads for different appliances. This can expand the load dataset and facilitate the process of load monitoring and prediction, etc.

Besides applications on load monitoring and reconstruction, load signatures can be used for other applications such as load forecasting and demand response. It provides unique labels to identify individual appliances. It is an important step to acquire the appliance operating information in a low-cost way without monitoring each appliance intrusively.

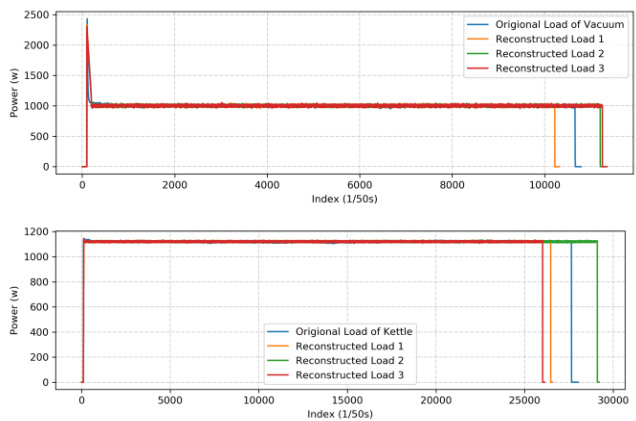

Fig. 18 Reconstructed load of Kettle (top) and Vacuum (bottom).

## V. Conclusion

This paper proposes a dynamic time-window event detection approach, the WAMMA method, for high-resolution data. Compared with existing event detection methods, the proposed method adjusts the parameters adaptively according to the data being processed. It overcomes the challenges such as high fluctuation, long transition, and near simultaneity, and greatly improves the event detection accuracy.

After event detection, representative transient and steady-state signatures are extracted and a sequential load signature tree representation is proposed as well to store load signatures. These signatures increase the identification accuracy of NILM and generate more realistic load data in a low-cost way.